\documentstyle[12pt]{article}
\def\al{\alpha}

\def\ga{\gamma}
\def\de{\delta}
\def\ep{\epsilon}

\def\et{\eta}

\def\si{\sigma}

\def\ph{\phi}

\def\ps{\psi}

\def\Ga{\Gamma}
\def\De{\Delta}

\def\Up{\Upsilon}

\def\fr#1#2{{{#1} \over {#2}}}

\def\bra#1{\langle{#1}|}
\def\ket#1{|{#1}\rangle}

\def\expect#1{\langle{#1}\rangle}

\def\half{{\textstyle{1\over 2}}}
\def\frac#1#2{{\textstyle{{#1}\over {#2}}}}

\def\lsim{\mathrel{\rlap{\lower4pt\hbox{\hskip1pt$\sim$}}
    \raise1pt\hbox{$<$}}}
\def\gsim{\mathrel{\rlap{\lower4pt\hbox{\hskip1pt$\sim$}}
    \raise1pt\hbox{$>$}}}
\def\sqr#1#2{{\vcenter{\vbox{\hrule height.#2pt
         \hbox{\vrule width.#2pt height#1pt \kern#1pt
         \vrule width.#2pt}
         \hrule height.#2pt}}}}

\def\Re{\hbox{Re}\,}
\def\Im{\hbox{Im}\,}

\def\z{{\bf\hat z}}

\newcommand{\beq}{\begin{equation}}
\newcommand{\eeq}{\end{equation}}
\newcommand{\bea}{\begin{eqnarray}}
\newcommand{\eea}{\end{eqnarray}}
\newcommand{\rf}[1]{(\ref{#1})}

\renewenvironment{thebibliography}[1]
 { \rm
   \begin{list}{\arabic{enumi}.}
    {\usecounter{enumi} \setlength{\parsep}{0pt}
     \setlength{\itemsep}{3pt} \settowidth{\labelwidth}{#1.}
     \sloppy
    }}{\end{list}}

\begin{document}
\titlepage

\begin{flushright}
{IUHET 285\\}
{hep-ph/9501372\\}
{October 1994\\}
\end{flushright}
\vglue 1cm

\begin{center}
{{\bf TESTS OF DIRECT AND INDIRECT CPT VIOLATION
\\}
{\bf AT A B FACTORY
\\}
\vglue 1.0cm
{Don Colladay and V. Alan Kosteleck\'y\\}
\bigskip
{\it Physics Department\\}
\medskip
{\it Indiana University\\}
\medskip
{\it Bloomington, IN 47405, U.S.A.\\}

\vglue 0.8cm
}
\vglue 0.3cm

\end{center}

{\rightskip=3pc\leftskip=3pc\noindent
The issue of testing CPT invariance at a $B$ factory
is considered.
We present asymmetries that permit a clean extraction
of quantities parametrizing direct and indirect CPT violation,
using information from $\Up(4S)$ decay
via coherent $B_d\overline {B_d}$ pairs
into various final states.
Estimates are given of the bounds on CPT violation
that could be obtained in present and planned machines.

}

\vskip 1truein
\centerline{\it Accepted for publication in Physics Letters B}

\vfill
\newpage

\baselineskip=20pt

{\it\noindent 1. Introduction.}
Invariance under the combined discrete symmetry
CPT is known to be a feature of
local relativistic point-particle field theories
\cite{s1,s2,s3,s4,s5,s6,s7},
including the standard model.
The current bound on CPT violation is obtained from
experiments in the kaon system
\cite{c1,c2},
where one figure of merit is a few parts in $10^{18}$.
The basic reason underlying the feasibility of
high precision tests of CPT
is the interferometric nature of the kaon system,
in which a small mass difference and the possibility of
strangeness oscillations amplify effects that would
otherwise be unobservable.

The $B^0$-$\overline{B^0}$ system is also
an interferometric system,
but one with detailed properties different from the kaon system.
In addition to its use as a probe
of standard-model features and parameters,
it is also possible at least in principle
to use a $B$ factory to further test CPT invariance.
At present, several efforts are underway
to develop $B$ factories,
for example,
at Cornell and SLAC in the U.S.
and at KEK in Japan.\footnote{
A discussion of some of the
many extant proposals can be found in ref.\
\cite{bf93}.
}
These machines are
designed to create relatively high fluxes of $\Up(4S)$
and hence of correlated
$B^0_d$-$\overline{B^0_d}$ pairs.

Other than intrinsic interest,
additional motivation for testing CPT comes from the possibility
that it is violated in the context of string theory,
ultimately due to the nonlocal nature of strings
\cite{kp1}.
It is natural to expect any effects of this type to be
strongly suppressed at accessible energy scales.
However,
the interferometric sensitivity of neutral-meson systems
may make such effects observable
\cite{kp2,kp3}.
In any neutral-meson system,
the string scenario suggests that
direct CPT violation would be too small to measure,
so all complex direct-violation parameters are effectively zero.
In contrast,
for each meson system the real and imaginary parts of
the complex indirect-violation parameter
satisfy a certain condition (see Eq.\ \rf{iib} below),
while the magnitude of this parameter
can in principle take values attainable
in the present or next generation of machines.

Experimentally testing these ideas
evidently requires isolating the various quantities
parametrizing CPT violation.
In the decay of a vector meson to a correlated neutral-meson pair,
a CP-violating effect
(together with the associated T or CPT violation)
can be extracted
through consideration of selected asymmetries in decay rates
to various final states.
For the $B$ system,
the literature contains some asymmetries
that vanish if CPT is preserved
\cite{kp3,ks,zx}.
However,
the issue of a clean extraction of CPT-violation parameters
has been an open problem because
T and CPT effects have not yet been disentangled.
One purpose of this paper is to fill this gap.
We present a means of separating quantities
parametrizing both direct and indirect CPT violation
by identifying certain asymmetries that isolate these parameters.
We also obtain estimates for the bounds on CPT violation
that could be obtained in present and planned $B$ factories
using these asymmetries.

\vglue 0.6cm
{\it \noindent 2. Preliminaries.}
In this section,
we present key definitions and equations
used later in the paper.
Generally,
our conventions
\cite{kp3}
are analogues of standard ones in
the kaon system.
Throughout,
we take all CP violation
(and hence T and CPT violation) to be small,
and we neglect terms that are higher-order
in small quantities.

The eigenvectors of the effective hamiltonian
for the $B^0$-$\overline{B^0}$ system
are given by
\bea
\ket{B_S} & = & \frac 1 {\sqrt 2}
[(1 + \ep_B + \de_B)\ket{B^0}
+(1 - \ep_B - \de_B)\ket{\overline{B^0}}]
\quad , \nonumber\\
\ket{B_L} & = & \frac 1 {\sqrt 2}
[(1 + \ep_B - \de_B)\ket{B^0}
-(1 - \ep_B + \de_B)\ket{\overline{B^0}}]
\quad .
\label{iia}
\eea
The CP-violating complex parameters $\ep_B$ and $\de_B$ are measures
of indirect T and indirect CPT violation, respectively.
For completeness,
we note here that in the string-inspired scenario for CPT violation
the quantity $\de_B$ satisfies
\beq
\Im\de_B = \pm\fr {\De\ga}{2\De m} \Re\de_B
\quad ,
\label{iib}
\eeq
where
$\De \ga = \ga_S - \ga_L$ and $\De m = m_L - m_S$
denote lifetime and mass differences, respectively,
of the physical particles $B_S$ and $B_L$.
Except where stated,
we do \it not \rm impose the condition \rf{iib} in what follows.

The initial state $\ket{i}$ of the
$B_d$-$\overline{B_d}$ pair arising
from the $\Up(4S)$ decay has $J^{PC}=1^{--}$.
Taking the direction of the $B$-meson momenta
to be along the $z$ axis,
this state can be written as
\beq
\ket{i} = \frac 1 {\sqrt 2}
[\ket{B_S(\z) B_L(-\z)} - \ket{B_L(\z) B_S(-\z)}]
\quad ,
\label{iic}
\eeq
where the argument $(\pm\z )$ denotes a meson
moving in the $\pm\z$ direction.
In what follows,
we label the two mesons by $\al = 1,2$
and take them to decay into final states $\ket{f_\al}$
at times $t_\al$
as measured in the rest frame of the $\Up(4S)$ decay.
Defining the transition amplitudes
\beq
a_{\al S} = \bra{f_\al}T\ket{B_S}~~,~~~~
a_{\al L} = \bra{f_\al}T\ket{B_L}
\quad
\label{iid}
\eeq
and their ratios
$\et_\al = a_{\al L}/a_{\al S}$,
it follows that the amplitude
${\cal{A}}_{12}(t_1, t_2)$ for the decay is
\beq
{\cal{A}}_{12}(t_1, t_2)
=\frac 1 {\sqrt 2} a_{1S}a_{2S}
\bigl (
\eta_2e^{-i(m_St_1+m_Lt_2)-\half (\ga_St_1+\ga_Lt_2)}
-\eta_1e^{-i(m_Lt_1+m_St_2)-\half (\ga_Lt_1+\ga_St_2)}
\bigr ) .
\label{iida}
\eeq

Experiments observe integrated rates.
It is useful to introduce first
the once-integrated decay rate
\beq
I(f_1,f_2,\pm v) = \half
\int_{v}^\infty dt~
|{\cal{A}}_{12}(t_1, t_2)|^2
\quad ,
\label{iidb}
\eeq
where $t=t_1+t_2$ is the sum of the decay times
and $v = |t_{2} - t_{1}|$
is the magnitude of their difference.
Calculation gives
\cite{kp3}
\beq
I(f_1,f_2,\pm v)
= \fr{|a_{1S}a_{2S}\et_{1}|^{2}} {2\ga}
e^{-\half \ga v}
\Bigl[ e^{\mp \frac 1 {2} \De \ga v} +
|r_{21}|^{2}e^{\pm\frac 1 {2} \De \ga v}
- 2|r_{21}| \cos (\De m v \pm \De \ph)\Bigr]
\quad ,
\label{iie}
\eeq
where
$\ga = \ga_S + \ga_L$,
$r_{21} = \et_2/\et_1$,
and $\De \ph = \ph_1 - \ph_2$,
with $\ph_\al$ given by
$\et_\al \equiv |\et_\al | e^{i\ph_\al}$.
It is also useful to define
symbols for two frequently occurring combinations
of the basic parameters $\De m$, $\ga$, and $\De\ga$:
\beq
a^2 = \De m^2 + \fr 14 \De \ga^2 ~~,~~~~
b^2 = \De m^2 + \fr 14 \ga^2
\quad .
\label{ab}
\eeq

In subsequent sections,
we use the following twice-integrated decay rates:
\bea
\Ga(f_1,f_2) & = & \int_{-\infty}^{\infty} dv
\ I(f_1,f_2,v) \nonumber \\
& = & \fr 1 {2\ga_S\ga_L}
\Bigl[ |a_{1S}a_{2L}|^2 + |a_{1L}a_{2S}|^2 - \fr{\ga_S\ga_L}{b^2}
(a^*_{1S}a^*_{2L}a_{1L}a_{2S} + {\rm c.c.})\Bigr]
\quad ,
\label{rate} \\
\Ga^+(f_1,f_2) & = & \int_0^\infty dv
\ I(f_1,f_2,v) \nonumber \\
& = & \fr 1{2\ga} \Bigl[
\fr{|a_{1S}a_{2L}|^2}{\ga_L} + \fr{|a_{1L}a_{2S}|^2}{\ga_S}
- (\fr{a^*_{1S}a^*_{2L} a_{1L}a_{2S}} {\half\ga - i \De m} + {\rm c.c.})
\Bigr]
\quad ,
\label{ratep} \\
\Ga^-(f_1,f_2) & = & \int_{-\infty}^0 dv
\ I(f_1,f_2,v) \nonumber \\
& = & \fr 1{2\ga} \Bigl[
\fr{|a_{1S}a_{2L}|^2}{\ga_S} + \fr{|a_{1L}a_{2S}|^2}{\ga_L}
- (\fr{a^*_{1S}a^*_{2L}a_{1L}a_{2S}}{\half\ga + i \De m} + {\rm c.c.})
\Bigr]
\quad ,
\label{ratem} \\
\Ga^+_{\rm incl}(f_1) & = & \sum_{f_2}\Ga^+(f_1,f_2)
\nonumber \\
& = & \fr 1{2\ga} \Bigl[ |a_{1S}|^2 + |a_{1L}|^2
- 2[a^*_{1S} a_{1L}(\Re \ep_B + i \Im \de_B) + {\rm c.c.}]
\Bigr]
\quad .
\label{incl}
\eea
Note that we do \it not \rm need the quantity
$\Ga^-_{\rm incl}(f_1)
\equiv \sum_{f_2}\Ga^-(f_1,f_2) \ne \Ga^+_{\rm incl}(f_1)$
in what follows.

\vglue 0.6cm
{\it\noindent 3. Direct CPT Violation.}
In this section we consider certain special $B$ decays,
called \it semileptonic-type \rm decays,
for each of which we present an asymmetry that extracts
a corresponding parameter measuring direct CPT violation.
These decays include the usual semileptonic decays,
along with a special class of other modes $B^0 \rightarrow f$
for which there is no lowest-order weak process that would allow
a significant contamination of either $\overline{B^0} \rightarrow f$
or $B^0 \rightarrow \overline f$.
Among the decays observed to date
\cite{pdt},
the semileptonic-type ones include
$B^0\rightarrow D^-D_s^+$,
$B^0\rightarrow J/\ps K^+\pi^-$,
$B^0\rightarrow J/\ps K^{*0}(892)$,
$B^0\rightarrow \ps(2S)K^{*0}(892)$,
and similar decays into excited states.
Note that the mode
$B^0\rightarrow J/\ps K^0$ is excluded
because the $K^0$ is not a directly observable final state.
For the other modes that have been seen,
a CKM-suppressed process exists contributing
to the contaminating transitions.

The various transition amplitudes
associated with the decay of the neutral $B$ meson
to a semileptonic-type final state $f$
can be parametrized as follows
\cite{lw,td}:
\bea
\bra{f}T\ket{B^0} = F_f (1 - y_f)~~~~,~~ &
\bra{f}T\ket{\overline{B^0}} = x_f F_f (1 - y_f)
\quad , \nonumber \\
\bra {\overline f}T\ket {\overline{B^0}} =
F_f^*(1 + y_f^*)~~~~,~~ &
\bra {\overline f}T\ket{B^0} = {\overline x_f^*} F_f^* (1 + y_f^*)
\quad .
\label{iiia}
\eea
In these expressions,
the parameters on the right-hand side are all complex.
The independent complex quantities
$x_f$ and $\overline x_f$ are included
to allow for the possibility of a violation in the
$\De B = \De Q$ rule.
They vanish if the rule is exact,
so in what follows we treat them as small.
If T invariance holds,
all the quantities $x_f$, $\overline x_f$, $F_f$, and $y_f$ are real.
If CPT invariance holds,
$x_f = \overline x_f$
and $y_f = 0$.
The parameter $y_f$ is therefore a measure
of direct CPT violation
in the decay to the final state $f$.
Its real part $\Re y_f$
is the present focus of our attention.

Each individual final state $f$ offers
the opportunity to test direct CPT violation
through a measurement of the corresponding $\Re y_f$.
To extract $\Re y_f$ from rate information,
we first determine the amplitudes
introduced in Eq.\ \rf{iid}.
Using the above definitions,
to first order in small quantities we find
\bea
a_{fS} & = & \frac 1 {\sqrt 2} F_f
(1 + \ep_B + \de_B - y_f + x_f)
\quad , \nonumber \\
a_{fL} & = & \frac 1 {\sqrt 2} F_f
(1 + \ep_B - \de_B - y_f - x_f)
\quad , \nonumber \\
a_{{\overline f}S} & = & \frac 1 {\sqrt 2} F^*_f
(1 - \ep_B - \de_B + y^*_f + \overline x_f^*)
\quad , \nonumber \\
a_{{\overline f}L} & = & - \frac 1 {\sqrt 2} F^*_f
(1 - \ep_B + \de_B + y^*_f - \overline x_f^*)
\quad .
\label{iiib}
\eea

With these amplitudes,
we can calculate
for the correlated $B$ pairs
the inclusive decay rates
$\Ga^+_{\rm incl}(f)$ of the type in Eq.\ \rf{incl},
where one final state
is required to be $f$ while the other is arbitrary
and where the decay into $f$ occurs first.
We can then extract the asymmetry $A^+_f$
between decays into $f$ and $\overline f$.
This asymmetry is proportional to $\Re y_f$.

Explicit calculation gives\footnote{
An analysis keeping terms to all orders
in $x_f$ and $\overline x_f$ shows that
Eq.\ \rf{a} has no linear corrections in these quantities.}
\beq
A^+_f \equiv \fr
{\Ga^+_{\rm incl}(f) - \Ga^+_{\rm incl}(\overline f)}
{\Ga^+_{\rm incl}(f) + \Ga^+_{\rm incl}(\overline f)}
= -2 \Re y_f
\quad .
\label{a}
\eeq
This asymmetry provides a clean way of extracting
an effect from direct CPT violation,
independently of any T or indirect CPT violation.
In section 5 below,
we comment on the experimental feasibility of using this asymmetry
and we provide an estimate of the bound attainable
on $\Re y_f$ using Eq.\ \rf{a}.

Although not central to the purpose of the present paper
it is worth noting that,
once a bound (or value) on $\Re y_f$ has been extracted,
the quantity $\Re\ep_B$ measuring T violation
can be determined without making the assumption of CPT invariance.
Consider the total integrated rate asymmetry
$A_{f,\overline f}^{\rm tot}$,
given by
\beq
A_{f,\overline f}^{\rm tot} \equiv \fr
{\Ga(f,f) - \Ga(\overline f, \overline f)}
{\Ga(f,f) + \Ga(\overline f, \overline f)}
= 4 \Re (\ep_B - y_f)
\quad .
\label{b}
\eeq
We see that a nonzero value of the combination
$\frac 14(A_{f,\overline f}^{\rm tot} - 2 A^+_f)\equiv\Re\ep_B$
for any given final state $f$ of the semileptonic type
can be unambiguously
attributed to the T-violation parameter $\Re \ep_B$
without making the assumption of CPT invariance.

We also note that the derivation of
Eqs.\ \rf{a} and \rf{b}
applies also to the $K^0$-$\overline{K^0}$ system,
when the final state is $f = \pi^- l^+ \nu_l$.
The quantities $\Re y_l$ and $\Re\ep_K$ for this system
can therefore also be obtained in this way.
Indeed,
one can show that,
in the absence of CPT invariance
and without resorting to a fit to a $\De t$-dependent quantity,
this method is the \it only \rm way to extract $\Re\ep_K$
from integrated asymmetries in $\ph$ decay.

\vglue 0.6cm
{\it \noindent 4. Indirect CPT violation.}
The complex parameter $\de_B$
is a measure of indirect CPT violation.
We first consider a means of obtaining its real part
and subsequently address the issue of the imaginary part.

Consider decays of the correlated $B$ pair
into either $J/\ps K_S$ or $J/\ps K_L$ in one decay channel
and a semileptonic-type state $f$ in the other.
In analogy with Eq.\ \rf{iiib},
we define the transition amplitudes
to the states involving $K^0$ and $\overline {K^0}$
as follows:
\bea
\bra{J/\ps K^0}T\ket{B^0}
= F_{J/\ps}(1 - y_{J/\ps})~~~~, &
\bra{J/\ps K^0}T\ket{\overline{B^0}}
= x_{J/\ps} F_{J/\ps}(1 - y_{J/\ps})
\quad , \nonumber \\
\bra{J/\ps \overline{K^0}}T\ket{\overline{B^0}}
= F^*_{J/\ps}(1 + y^*_{J/\ps})~~~~, &
\bra{J/\ps \overline{K^0}}T\ket{B^0}
= \overline x^*_{J/\ps} F^*_{J/\ps}(1 + y^*_{J/\ps})
\quad .
\label{iva}
\eea
As before,
this allows for possible violation
of the $\De B = \De Q$ rule
via $x_{J/\ps}$ and $\overline x_{J/\ps}$,
while the complex parameter $y_{J/\ps}$
characterizes direct CPT violation.
These parameters are assumed small in what follows.

The final products of the $\Up(4S)$ decay involve $K_S$ and $K_L$
rather than $K^0$ and $\overline{K^0}$.
The ratios of matrix elements useful for asymmetry determination
therefore involve the former states.
Using the definitions \rf{iva},
we obtain:
\bea
\et_{J/\ps K_S} & \equiv &
\fr{\bra{J/\ps K_S}T\ket{B_L }}
{\bra{J/\ps K_S}T\ket{B_S}}
\nonumber \\
& = &
\ep_K^* + \ep_B + \de_K^* - \de_B
- \fr {\Re (F_{J/\ps} y_{J/\ps})} {\Re F_{J/\ps}}
- \fr 1 {2\Re F_{J/\ps}}
(x_{J/\ps} F_{J/\ps} - \overline x^*_{J/\ps} F^*_{J/\ps})
\nonumber \\
& & +~ i \fr {\Im F_{J/\ps}} {\Re F_{J/\ps}}
\Bigl[ 1
- i \fr {\Im F_{J/\ps}} {\Re F_{J/\ps}}
(\ep_K^* + \ep_B + \de_K^* + \de_B )
\nonumber \\
& & \qquad\qquad
- \fr 1 {2\Re F_{J/\ps}}
(x_{J/\ps} F_{J/\ps} + \overline x^*_{J/\ps} F^*_{J/\ps})
+ i \fr {\Im (F_{J/\ps} y_{J/\ps})} {\Re F_{J/\ps}}
\Bigr]
\quad ,
\nonumber \\
\et_{J/\ps K_L} & \equiv &
\fr{\bra{J/\ps K_L}T\ket{B_S}}
{\bra{J/\ps K_L}T\ket{B_L}}
\nonumber \\
& = &
\ep_K^* + \ep_B - \de_K^* + \de_B
- \fr {\Re (F_{J/\ps} y_{J/\ps})} {\Re F_{J/\ps}}
+ \fr 1 {2\Re F_{J/\ps}}
(x_{J/\ps} F_{J/\ps} - \overline x^*_{J/\ps} F^*_{J/\ps})
\nonumber \\
& &  +~  i \fr {\Im F_{J/\ps}} {\Re F_{J/\ps}}
\Bigl[ 1
- i \fr {\Im F_{J/\ps}} {\Re F_{J/\ps}}
(\ep_K^* + \ep_B - \de_K^* - \de_B )
\nonumber \\
& & \qquad\qquad
+ \fr 1 {2\Re F_{J/\ps}}
(x_{J/\ps} F_{J/\ps} + \overline x^*_{J/\ps} F^*_{J/\ps})
+ i \fr {\Im (F_{J/\ps} y_{J/\ps})} {\Re F_{J/\ps}}
\Bigr]
\quad .
\label{ivb}
\eea
In these equations,
the parameters $\ep_K$ and $\de_K$ are
quantities parametrizing indirect T and CPT violation
in the kaon system.

The goal is to identify an asymmetry or combination of asymmetries
permitting the extraction of $\Re \de_B$.
To this end,
we introduce the following two rate asymmetries:
\bea
A_{f,K_S} & \equiv &
\fr{\Ga(f,J/\ps K_S) - \Ga(\overline f,J/\ps K_S)}
{\Ga(f,J/\ps K_S) + \Ga(\overline f,J/\ps K_S)}
\nonumber \\
& = & 2\Re (\ep_B - y_f - \de_B)
- \fr{2\ga_S\ga_L}{b^2}\Re (\et_{J/\ps K_S})
\quad ,
\label{ivc}
\eea
and
\bea
A_{f,K_L} & \equiv &
\fr{\Ga(f,J/\ps K_L) - \Ga(\overline f,J/\ps K_L)}
{\Ga(f,J/\ps K_L) + \Ga(\overline f,J/\ps K_L)}
\nonumber \\
& = & 2\Re (\ep_B - y_f + \de_B)
- \fr{2\ga_S\ga_L}{b^2} \Re (\et_{J/\ps} K_L)
\quad .
\label{ivd}
\eea
In deriving the explicit form of these two asymmetries,
we have assumed that violations of the $\De B=\De Q$ rule
are independent of violations of CPT invariance,
so that
$x_f = \overline x_f$ and
$x_{J/\ps} = \overline x_{J/\ps}$.
Since $\Im F_{J/\ps}$ controls the direct T violation
in these processes,
we have also treated it as a small quantity.

The difference between the asymmetries
in Eqs.\ \rf{ivc} and \rf{ivd}
is a function of CPT-violating parameters:\footnote{
A derivation relaxing the constraint of small direct T violation
shows that Eq.\ \rf{ive} is correct up to terms
simultaneously quadratic in
$\Im F_{J/\ps}$ and linear in $\de_B$ or $\de_K$.}
\beq
A_{L,S} \equiv
A_{f,K_L} - A_{f,K_S} =
\fr 4{b^2}[a^2 \Re \de_B
+ \ga_S\ga_L \Re \de_K]
\quad .
\label{ive}
\eeq
The measurement of $A_{L,S}$ provides
a means of obtaining a fairly stringent
bound on $\Re\de_B$.
The point is that the parameter $\Re\de_K$ can be bounded
using rate information from semileptonic $K$ decays.
This is discussed further in the next section.
Note also that this result is independent of the state $f$,
which makes the statistics more favorable by allowing a
sum over the class of semileptonic-type final states.
We emphasize that the asymmetry combination $A_{L,S}$
permits the extraction of $\Re \de_B$
independently of effects from direct CPT violation
and direct or indirect T violation
arising in either the $B$ or the $K$ systems.

Once a bound on $\Re\de_B$ is established,
$\Im\de_B$ can in turn be obtained from a measurement
of double semileptonic decay rates of the $\Up(4S)$.
The quantity to be measured is
\cite{kp3}
\beq
A_{f,\overline f} \equiv
\fr {\Ga^+(f,\overline f) - \Ga^-(f,\overline f)}
{\Ga^+(f,\overline f) + \Ga^-(f,\overline f)} =
4 \fr {b^2 \De \ga \Re\de_B + 2 \De m \ga_S\ga_L \Im\de_B}
{\ga(b^2 + \ga_S\ga_L)}
\quad .
\label{ivf}
\eeq
In obtaining the explicit form of this asymmetry,
it is assumed that any violation of the $\De B=\De Q$ rule
is independent of CP violation,
so that $x_f = \overline x^*_f$.

\vglue 0.6cm
{\it\noindent 5. Estimates of bounds attainable.}
In this section,
we investigate the bounds attainable
from the above analyses
on the quantities
parametrizing direct and indirect CPT violation
in the $B$ system.
Following the methods of ref.\ \cite{dhr},
for each of the relevant quantities
we provide an estimate of
the number of $\Up (4S)$ events required to
reduce the error in
the associated asymmetry to one standard deviation.

In general,
for an asymmetry
$A = (N_+ - N_-)/(N_+ + N_-)$,
the binomial distribution implies
that the expected number of events $\expect{N_+}$
required to observe a nonzero $\expect{A}$
at the $N\si$ level is
$N^2(1 + \expect{A})(1 - \expect{A}^2)/2\expect{A}^2$.
To convert this to $\Up(4S)$ events,
this number must be multiplied
by two to account for
the branching ratio of $\Up (4S)$ into two neutral $B$ mesons
and by the inverse branching ratio for the latter
into the relevant final states.
The assumption that any T and CPT violations
are small implies that interference effects
in the correlated decays can be neglected.

We first consider bounds on the various $\Re y_f$,
which provide measures of direct CPT violation.
The relevant asymmetry is $A^+_f$,
given by Eq.\ \rf{a}.
Since the second final state is unrestricted,
it is sufficient to multiply only by the inverse branching ratio
for the process $B^0 \rightarrow f$.
An additional multiplicative factor
appears because the asymmetry involves only those
events for which the decay into $f$ occurs first.
This factor is two because in the
$B$ system $\ga_S\approx\ga_L$,
which makes $\De t > 0 $ events
about as likely as $\De t < 0 $ events.
Combining this information,
we find that the number
$N_{\Up (4S)}(\Re y_f)$ of $\Up(4S)$ events needed to
reduce the error in
$\Re y_f$ to within one standard deviation $\si$ is
\beq
N_{\Up (4S)}(\Re y_f) \simeq
\fr 1 {2\si^2{\rm BR}(B^0\rightarrow f)}
\quad .
\label{va}
\eeq

Next,
we consider the bound on $\Re\de_B$,
parametrizing indirect CPT violation.
For the combination $A_{L,S}$ of asymmetries
given by Eq.\ \rf{ive},
the errors in $A_{f,K_S}$ and $A_{f,K_L}$
must be combined in quadrature.
Also,
an estimate is needed of the size of the coefficients of
$\Re\de_B$ and $\Re\de_K$ in the equation.
For the latter,
we take
\cite{pdt}
$\ga_S \approx \ga_L$
and\footnote{
The value of $x$ quoted is a lower bound,
$|x|\geq 0.71$,
if CPT invariance is not assumed
\cite{ks}.
However,
a value above this bound improves the statistics
obtained below.}
$x =  2 \De m/\ga \simeq \pm 0.71$.
With these values,
Eq.\ \rf{ive} becomes
\beq
A_{L,S} \approx 1.3 \Re\de_B
+ 2.7 \Re\de_K
\quad .
\label{vb}
\eeq

Since $A_{L,S}$ is independent of the specific
semileptonic-type final state $f$,
the corresponding branching ratios can be summed.
This gives $\sum_f {\rm BR}(B^0 \rightarrow f) \simeq 15\%$.
Since the $K^0$ is roughly 50\% $K_S$ and 50\% $K_L$,
we take
\cite{pdt}
\beq
{\rm BR}(B^0\rightarrow J/\ps K_S) \approx
{\rm BR}(B^0\rightarrow J/\ps K_L)
\approx
\frac 12 {\rm BR}(B^0\rightarrow J/\ps K^0) \simeq
3.8 \times 10^{-4}
\quad .
\label{vc}
\eeq
{}From the limit cited in
ref.\ \cite{tr},
the current bound on $\Re\de_K$
lies at the $10^{-3}$ level.
For simplicity,
take $\Re\de_K$ to be zero,
i.e., sufficiently well bounded by $K$-decay experiments.
We also take the errors in the asymmetries
$A_{f,K_S}$ and $A_{f,K_L}$ to be roughly equal.
Then,
we find that the number
$N_{\Up (4S)}(\Re\de_B)$ of $\Up (4S)$ events
needed to reduce the error in $\Re\de_B$
to within one standard deviation $\si$ is
\beq
N_{\Up (4S)}(\Re\de_B) \simeq \fr {1.8\times 10^4} {\si^2}
\quad .
\label{vd}
\eeq

Finally,
given a bound on $\Re\de_B$,
Eq.\ \rf{ivf} can be used to
provide an estimate of the number
$N_{\Up (4S)}(\Im\de_B)$ of $\Up (4S)$ events
needed to reduce the error in $\Im\de_B$
to within one standard deviation $\si$.
For example,
if the string-inspired relation \rf{iib} is valid,
a similar calculation to those above gives
\cite{kp3}
\beq
N_{\Up (4S)}(\Im\de_B) \simeq \fr 5 {\si^2}
\quad .
\label{vf}
\eeq

The ease with which experimental information
can be obtained differs for the above quantities.
In particular,
the asymmetries involved in bounding $\Re y_f$ and $\Im\de_B$
require knowledge of the sign of $\De t$.
However,
at a symmetric $B$ factory
the distance between decay vertices of the two $B$ mesons
is only about $60\mu$m,
and information about the location of the
$\Up (4S)$ decay is difficult to acquire.
A symmetric $B$ factory is therefore best suited
to measure or bound $\Re\de_B$.
The situation is improved at an asymmetric $B$ factory,
where the boost alters the topology of the events
(see, for example,
ref.\ \cite{sch}.)
This creates a greatly decreased angular separation
and hence an easier determination of the sign of $\De t$.

\vglue 0.6cm
{\it\noindent 6. Summary.}
We have presented asymmetries
that allow the independent extraction of quantities
parametrizing both direct and indirect CPT violation
in the $B$ system.
These asymmetries are given
in Eq.\ \rf{a} for $\Re y_f$,
in Eq.\ \rf{ive} for $\Re \de_B$,
and in Eq.\ \rf{ivf} for $\Im \de_B$.
We have also shown that,
once direct violation is measured or bounded,
the quantity $\Re\ep_B$ parametrizing indirect T violation
can be obtained
from the asymmetry \rf{b} {\it without}
assumptions regarding CPT invariance.

Assuming no severe acceptance or background effects,
it appears experimentally feasible to put bounds on both
direct and indirect CPT violation.
Estimates of the bounds attainable
are given
in Eq.\ \rf{va} for $\Re y_f$,
in Eq.\ \rf{vd} for $\Re \de_B$,
and in Eq.\ \rf{vf} for $\Im \de_B$.
Bounding $\Re\de_B$ is possible at either a symmetric
or an antisymmetric $B$ factory.
This can be performed
by comparing $B_d$ decays
into $J/\ps K_S$ with decays into $J/\ps K_L$,
without the need for information about $\De t$.
Measurements of $\Re y_f$ and $\Im\de_B$
require a knowledge of the sign of $\De t$,
which is more easily obtained at an asymmetric factory.
Accumulation of about $10^7$ or $10^8$
correlated $B_d$-$\overline{B_d}$ pairs,
which could result from about one running year at
a $B$ factory meeting typical design luminosities,
should permit the determination of bounds on the various quantities
to approximately the $10^{-2}$ level.

Independent examination of the different possible types of
CPT violation in the $B$ system is worthwhile
since CPT invariance is a fundamental symmetry of the standard model.
If any violation is uncovered,
the possibility of stringy effects in the system can be tested
and the source can be isolated by the methods presented above.

\vglue 0.6cm
{\it\noindent Acknowledgment.}
This work was supported in part
by the United States Department of Energy
under grant number DE-FG02-91ER40661.

\vglue 0.6cm

\end{document}